\documentclass{iau}
\usepackage{graphicx}

\title[Discovery of SNRs in the $\lambda$6\ cm survey] 
{Discovery of supernova remnants in the Sino-German $\lambda$6\ cm
  polarization survey of the Galactic plane} 

\author[X. Y. Gao et al.]  
{X. Y.~Gao$^1$, X. H.~Sun$^1$, J. L.~Han$^1$, W.~Reich$^2$,
  P.~Reich$^2$ and R.~Wielebinski$^{2}$
}

\affiliation{1. National Astronomical Observatories, Chinese Academy
  of Sciences ,Jia-20 Datun Road, Chaoyang District, Beijing 100012,
  China.  bearwards@gmail.com \\ 2. Max-Planck-Institut f\"{u}r
  Radioastronomie, Auf dem H\"{u}gel 69, 53121 Bonn, Germany }

\pubyear{2013}
\volume{296}  
\pagerange{119--122}
\jname{\mbox{Supernova environmental impact}}
\editors{Dick McCray \& Alak Ray, ed.}
\begin{document}

\maketitle

\begin{abstract}
  The Sino-German $\lambda$6\ cm polarization survey has mapped in
  total intensity and polarization intensity over an area of
  approximately 2\,200 square degrees in the Galactic disk. This
  survey provides an opportunity to search for Galactic supernova
  remnants (SNRs) that were previously unknown. We discovered the new
  SNRs G178.2$-$4.2 and G25.1$-$2.3 which have non-thermal spectra,
  using the $\lambda$6\ cm data together with the observations with
  the Effelsberg telescope at $\lambda$11\ cm and
  $\lambda$21\ cm. Both G178.2$-$4.2 and G25.1$-$2.3 are faint and
  have an apparent diameter greater than 1$^{\circ}$. G178.2$-$4.2
  shows a polarized shell. HI data suggest that G25.1$-$2.3 might have
  a distance of about 3~kpc. The $\lambda$6\ cm survey data were also
  very important to identify two other new SNRs, G152.4$-$2.1 and
  G190.9$-$2.2.

  \keywords{Radio continuum: ISM -- ISM: supernova remnants --
    Polarization}
\end{abstract}

\firstsection 

\section{Introduction}

Supernova explosions have a substantial impact on the interstellar
environment. Supernova remnants (SNRs) are post-explosion relics, and
are formed when shocks from the explosion sweep up and interact with
the surrounding medium. Large-scale radio surveys are ideal hunting
grounds for new SNRs. In the most frequently used Galactic SNR
catalogue compiled by Dave Green (\cite{Green09}), mainly based on
radio continuum observations, there are 274 SNRs. \cite{Ferrand12}
recently made a new Galactic SNR catalogue by including new detections
from high energy observations. The number of known Galactic SNRs is
now 312. This quantity is still far less than the theoretical
predictions (e.g. \cite{Tammann94}), because of two major limitations:
the sensitivity and the angular resolution of the observations.

The Sino-German $\lambda$6\ cm polarization survey of the Galactic
plane (\cite{Sun07}, \cite{Gao10}, \cite{Sun11}, \cite{Xiao11}) was
conducted between the years 2004 and 2009, observing the Galactic disk
in the range of $10^{\circ} \leqslant l \leqslant 230^{\circ}$ and
$|b| \leqslant 5^{\circ}$. The angular resolution is 9.5$^{\prime}$,
and the average sensitivity (1$\sigma$ noise) of the survey is about
0.8~mK T$_{b}$ in total intensity $I$ and 0.5~mK T$_{b}$ in linear
polarization $U$ and $Q$. Although the angular resolution is coarser
in comparison with synthesis telescopes and large single dishes, the
system is more suitable for observing SNRs with large extent, and the
high sensitivity of the Sino-German survey enables us to discover SNRs
as faint as G156.2+5.7, the SNR with the lowest surface brightness
until recently.  One of the major goals of the $\lambda$6\ cm survey
is to study and identify Galactic SNRs (see Han et al. 2013, this
volume). In this talk, we present the discovery of two new SNRs
G178.2$-$4.2 and G25.1$-$2.3 in our $\lambda$6\ cm survey
(\cite{Gao11}).

\section{Identification of two new SNRs}

Considering the limitation in angular resolution, we search for
shell-type objects as SNR candidate. Shell-type SNRs often appear more
extended than the crab-like ones, and are easier to identify due to
three characteristics: 1) shell or partial shell structures, 2)
associated polarized emission within the shell, and 3) the non-thermal
spectrum with a spectral index around $\beta \sim$ $-$2.5 (S$_{\nu}
\sim \nu^{\beta}$, $\alpha = \beta + 2$), as expected for adiabatic
expansion with a compression factor of 4. These are the three criteria
for our SNR identifications. Note that polarization may not be
detected due to Faraday depolarization. We successfully identified two
new shell-type SNRs G178.2$-$4.2 and G25.1$-$2.3 in the $\lambda$6\ cm
survey.

\begin{figure*} 
\centering
\includegraphics[angle=-90, width=0.48\textwidth]{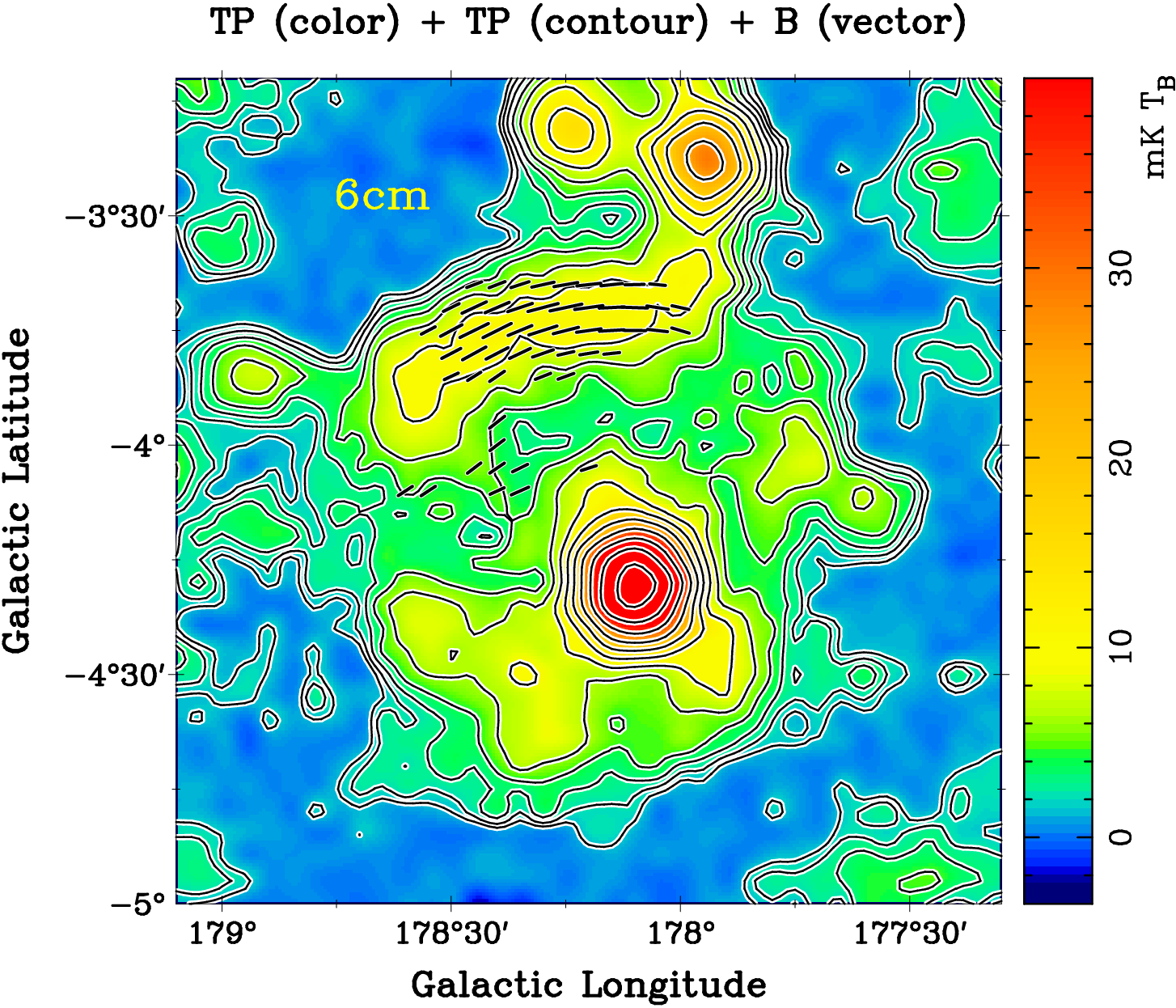}
\includegraphics[angle=-90, width=0.48\textwidth]{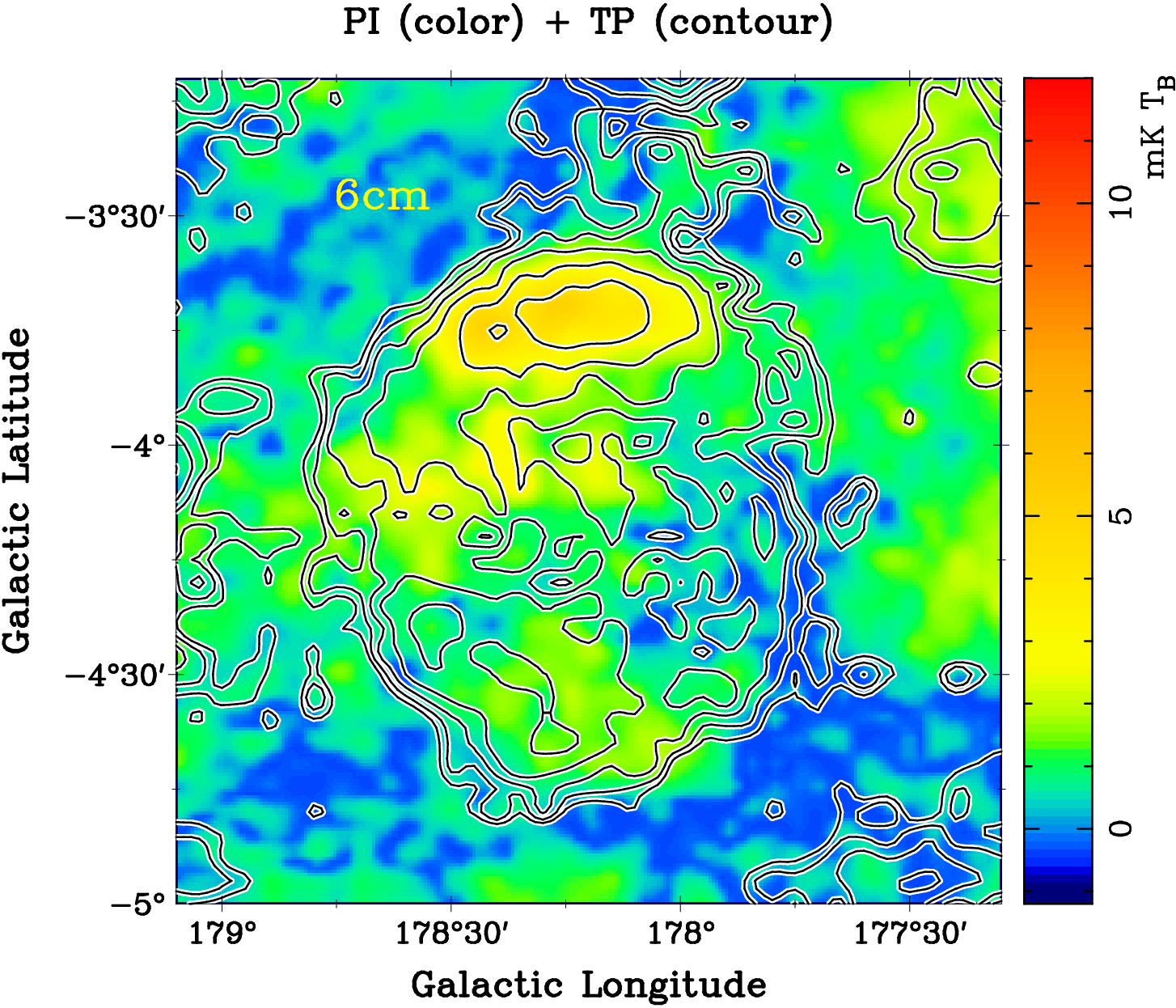}\\[2mm]
\includegraphics[angle=-90, width=0.48\textwidth]{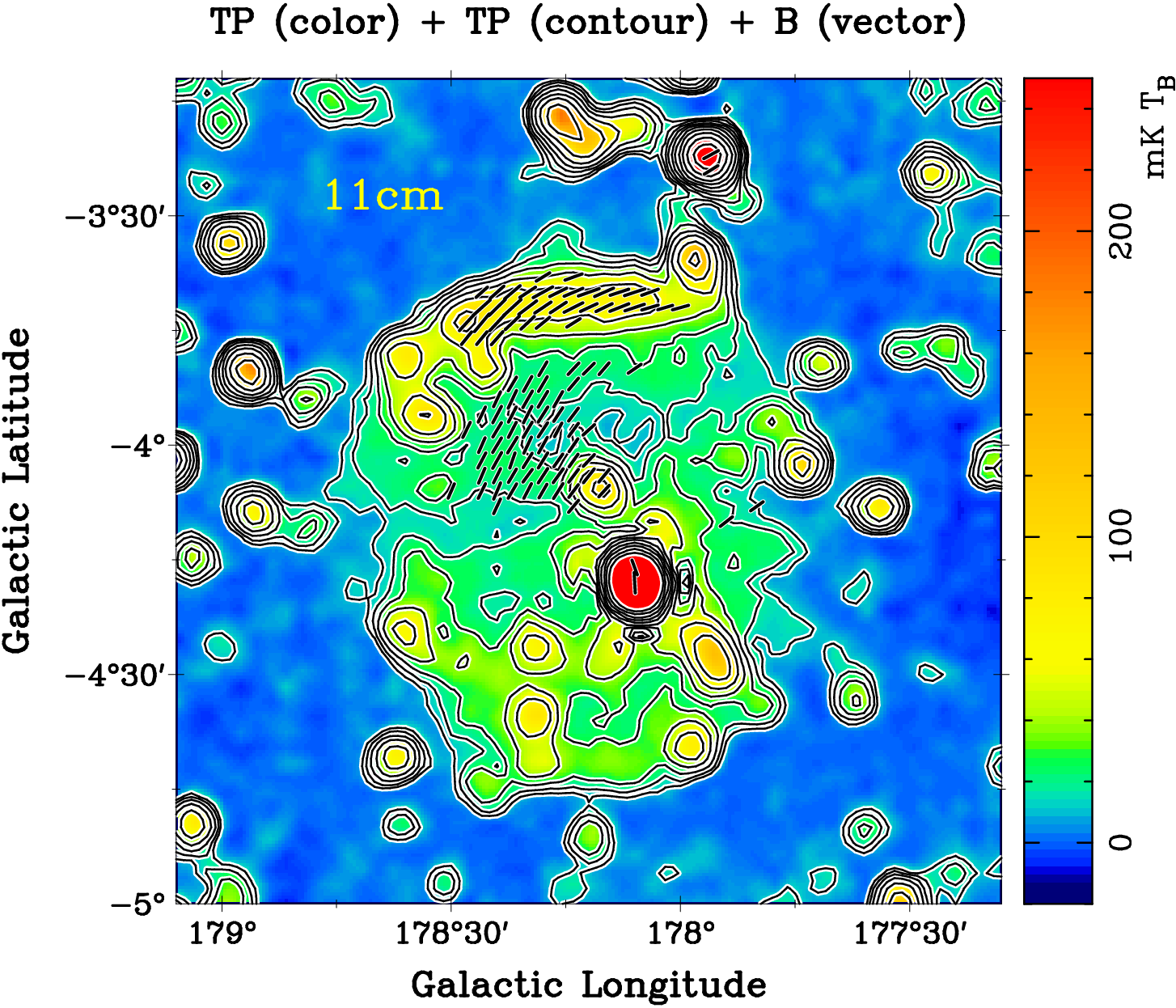}
\includegraphics[angle=-90, width=0.48\textwidth]{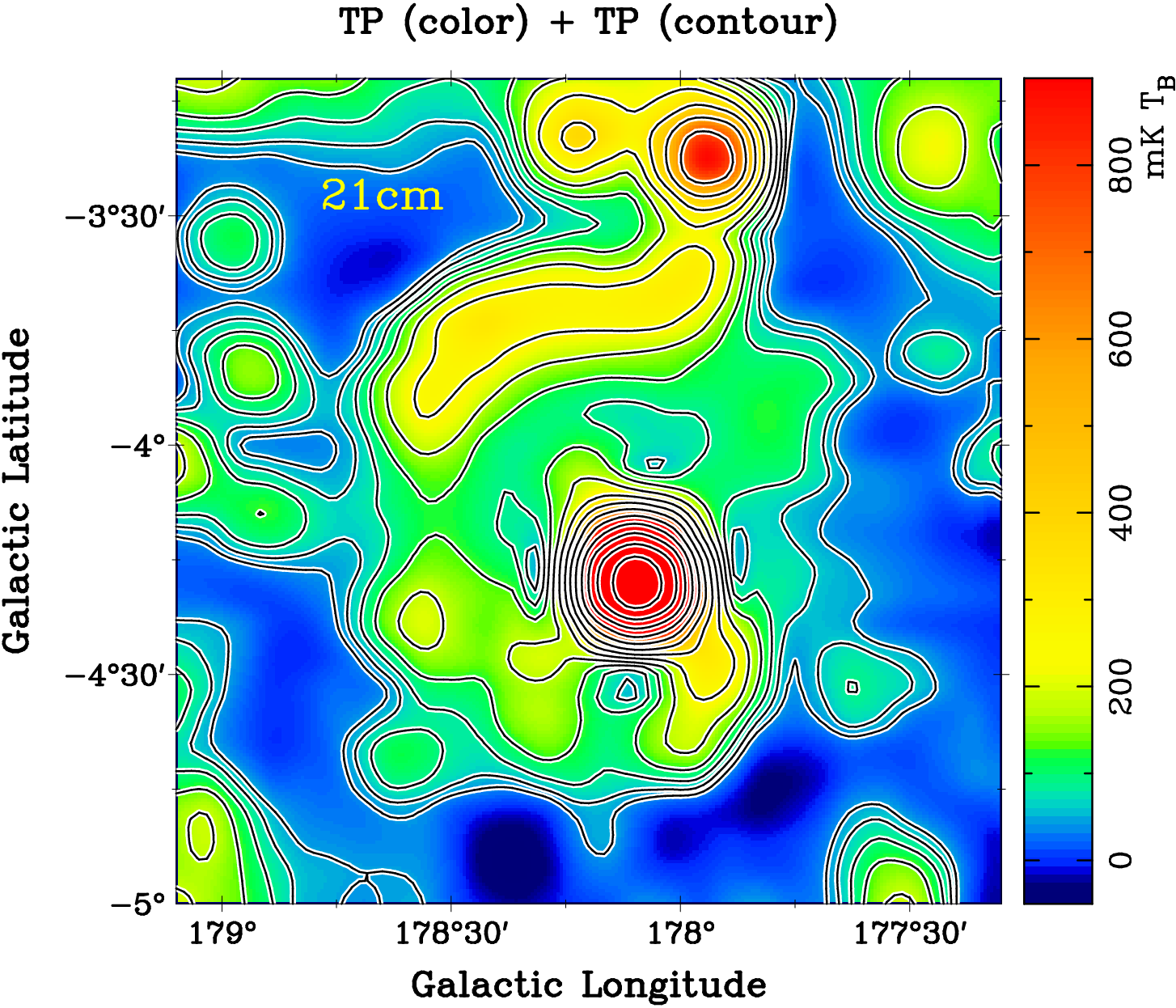}
\caption{Total intensity and polarization intensity images of the new
  SNR G178.2$-$4.2 measured at $\lambda$6\ cm, $\lambda$11\ cm and
  $\lambda$21\ cm. The top right panel shows the total intensity
  contours after subtracting point sources.}
\label{G178}
\end{figure*}

\begin{figure*} 
\centering
\includegraphics[angle=-90, width=0.45\textwidth]{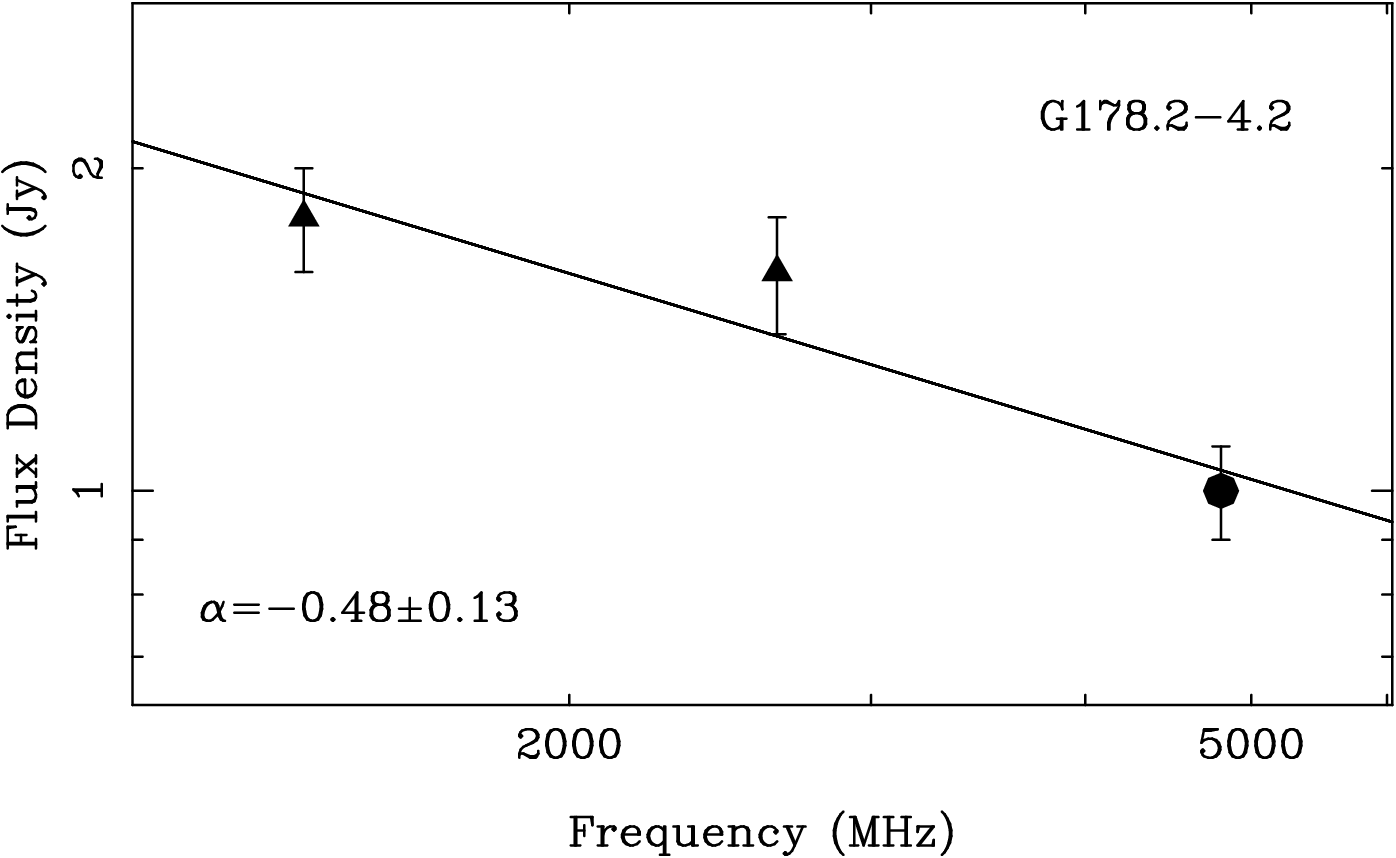}
\includegraphics[angle=-90, width=0.45\textwidth]{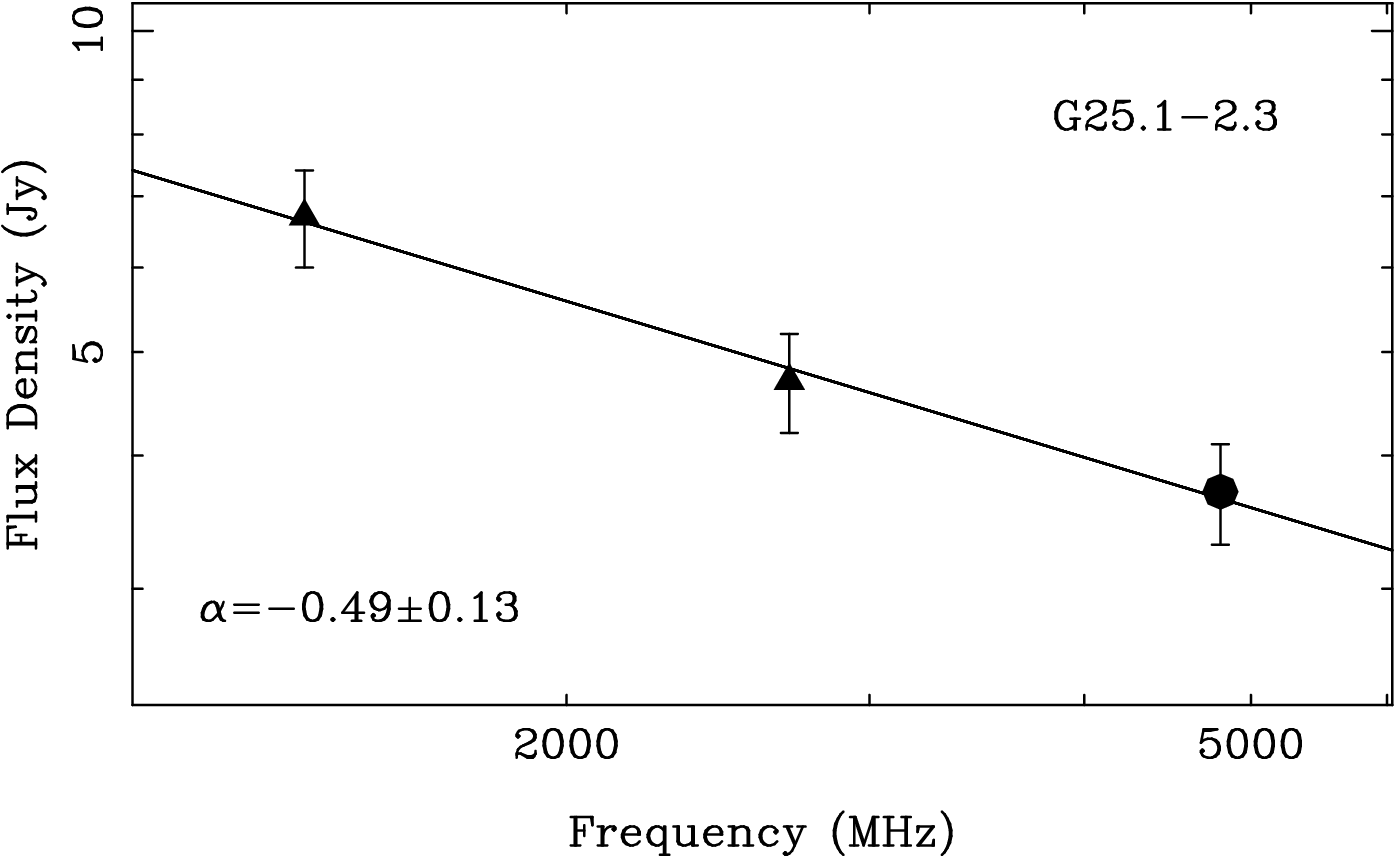}
\caption{Integrated radio spectrum of G178.2$-$4.2 and G25.1$-$2.3.}
\label{spe}
\end{figure*}

\begin{figure*} 
\includegraphics[angle=-90, width=0.48\textwidth]{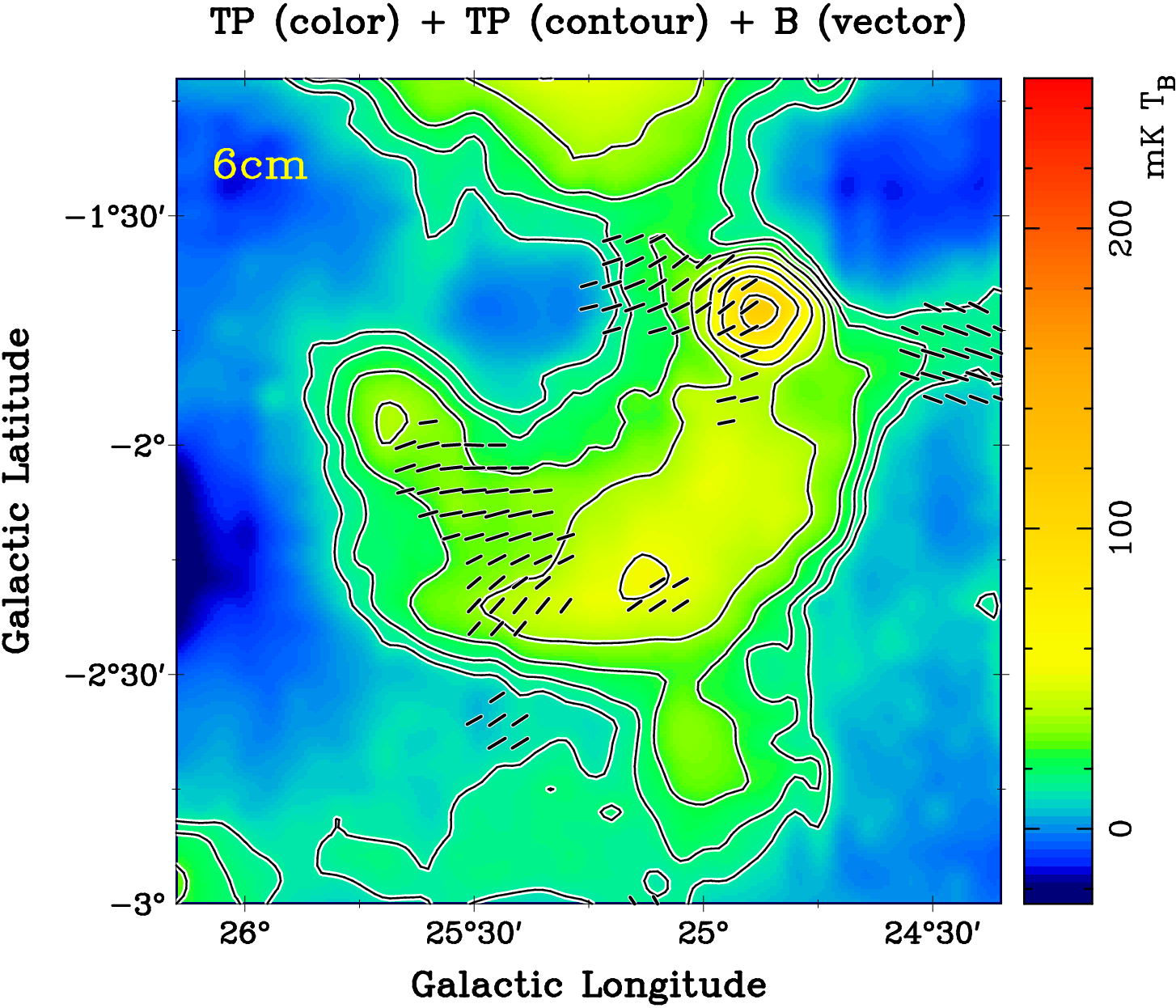}
\includegraphics[angle=-90, width=0.48\textwidth]{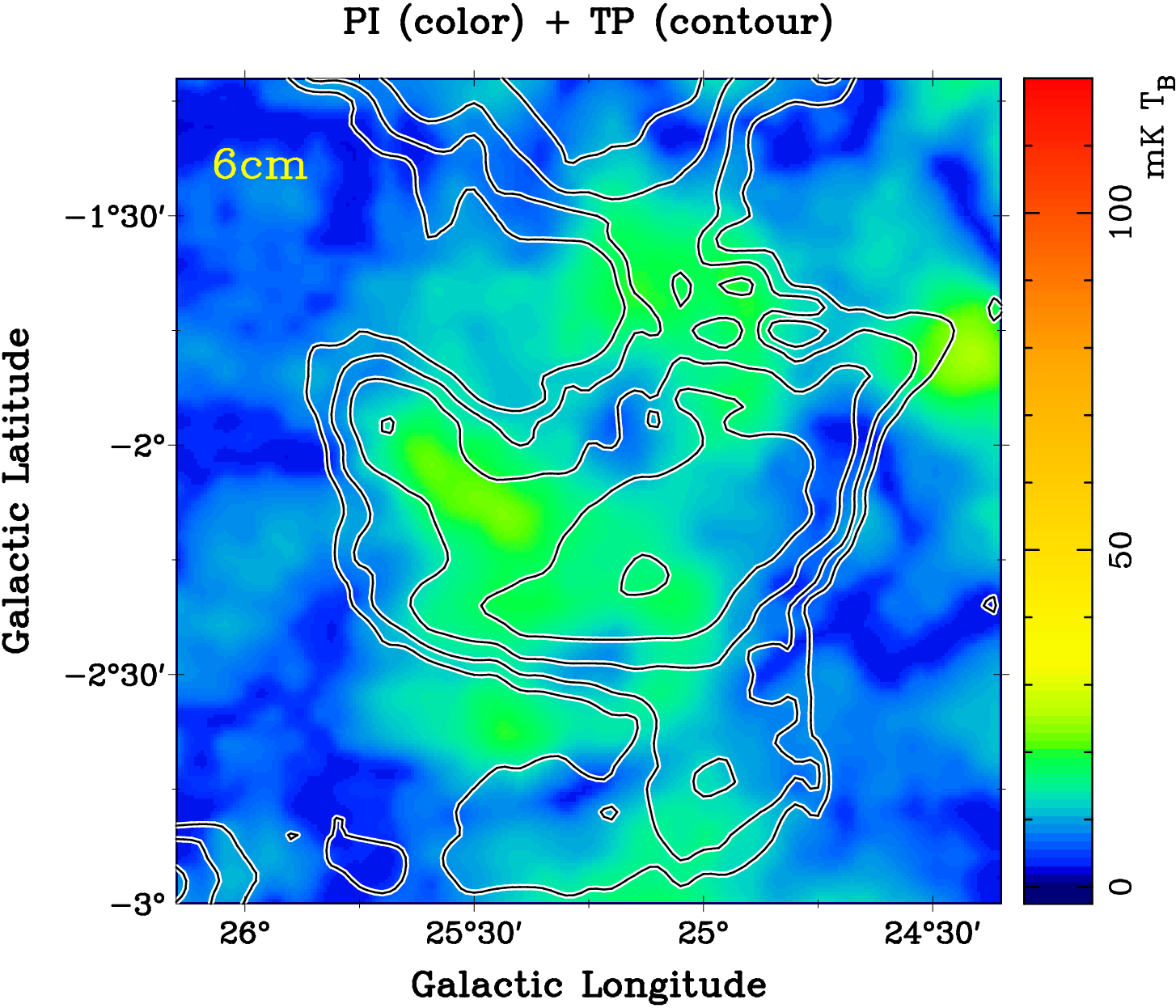}\\[2mm]
\includegraphics[angle=-90, width=0.48\textwidth]{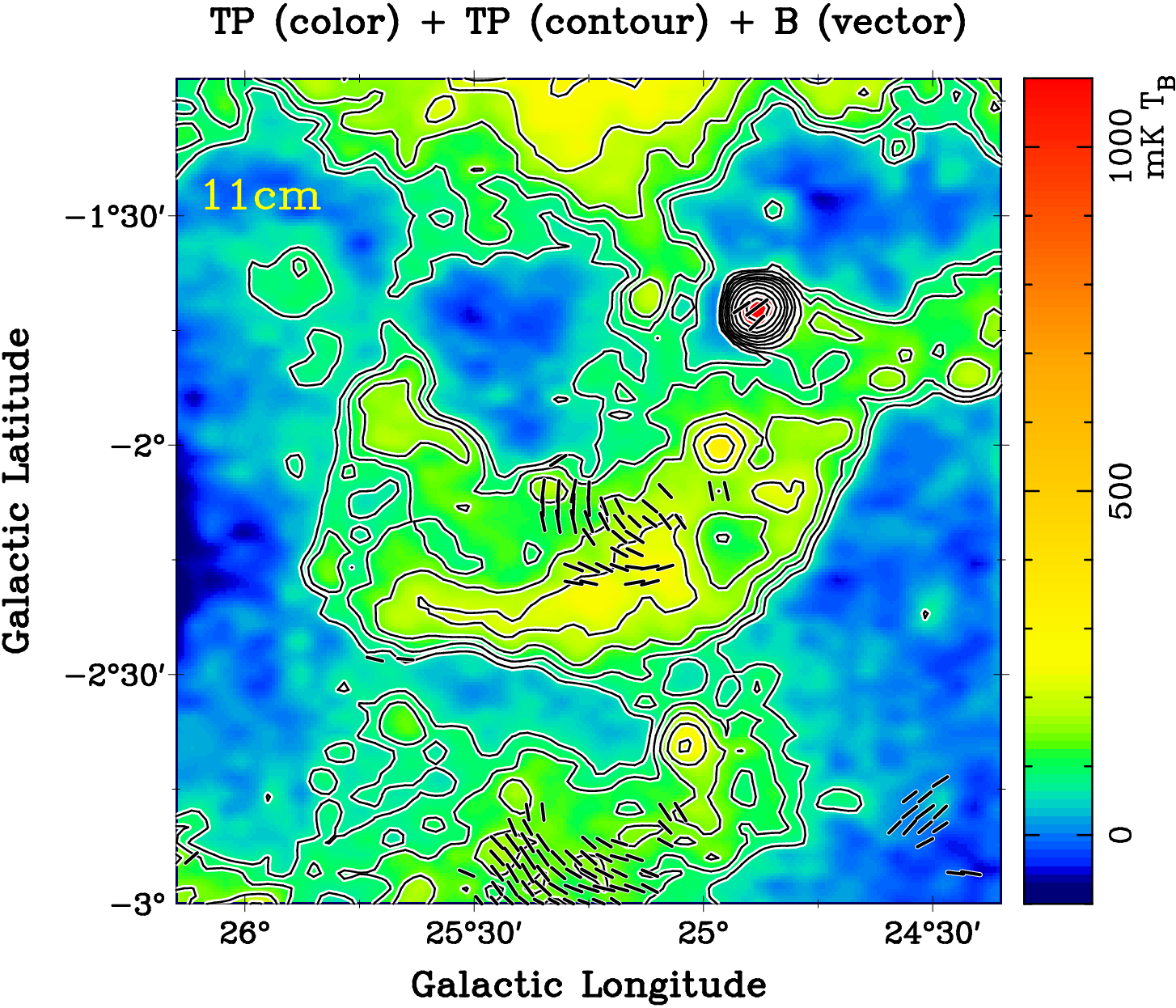}
\includegraphics[angle=-90, width=0.48\textwidth]{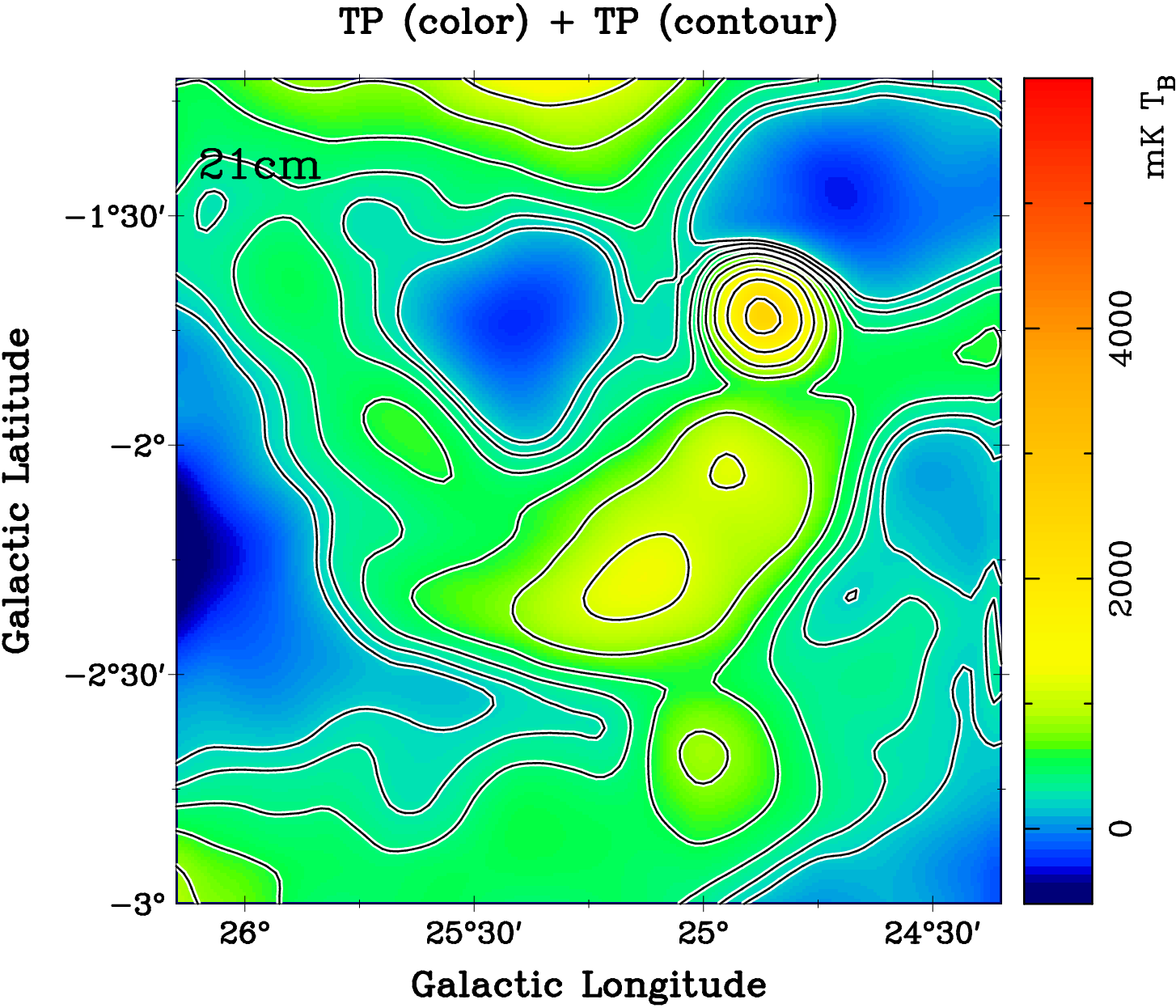}

\caption{The same as in Fig.~\ref{G178}, but for G25.1$-$2.3.}
\label{G25}
\end{figure*}

G178.2$-$4.2 is located in the anti-center region of the Galaxy
(Fig.~\ref{G178}). It has a circular shape with an apparent diameter
of around 1$^{\circ}$. A prominent shell is seen in its northern part.
The un-related, unresolved double-sided radio source 3C139.2 is near
the center of G178.2$-$4.2. Polarized emission is seen in the northern
shell at both $\lambda$6\ cm and $\lambda$11\ cm. B-field vectors
(\overrightarrow{E}+90$^{\circ}$) are found to be tangential within
the shell at $\lambda$6\ cm. We observed G178.2$-$4.2 at
$\lambda$11\ cm with the Effelsberg 100-m telescope in March, 2009,
and we extracted the $\lambda$21\ cm data from the Effelsberg
$\lambda$21\ cm survey of the Galactic plane (\cite{Reich97}) and the
Effelsberg $\lambda$21\ cm medium latitude survey
(\cite{Reich04}). The flux density is integrated over the same area of
G178.2$-$4.2 at $\lambda$6\ cm, $\lambda$11\ cm and $\lambda$21\ cm,
after removing the contribution from extra-Galactic sources and
background emission. We measured $S_{6cm}$ = 1.0$\pm$0.1~Jy,
$S_{11cm}$ = 1.6$\pm$0.2~Jy and $S_{21cm}$ = 1.8$\pm$0.2~Jy, yielding
an integrated spectral index of $\alpha = -0.48\pm0.13$
(Fig.~\ref{spe}). This value indicates the non-thermal nature of
G178.2$-$4.2. In summary, the shell structure, the polarized emission
and the non-thermal nature strongly indicate that G178.2$-$4.2 is a
SNR. From the integrated flux density, we calculated the surface
brightness of the new SNR G178.2$-$4.2 to be $\rm \Sigma_{1~GHz} = 7.2
\times 10^{-23}Wm^{-2}Hz^{-1}sr^{-1}$. This small value places it
among the faintest SNRs known in the Galaxy.

G25.1$-$2.3 is found in the inner part of the Galaxy. It is elusive
until we filter out the confusion from the strong diffuse Galactic
emission. We examined G25.1$-$2.3 using the data from the Urumqi
$\lambda$6\ cm survey, the Effelsberg $\lambda$11\ cm survey
(\cite{Reich9011}) and the Effelsberg $\lambda$21\ cm survey
(\cite{Reich9021}), and found that G25.1$-$2.3 has only one shell
curving to the south. Polarization patches are detected within the
shell at $\lambda$6\ cm and $\lambda$11\ cm, but they seem to be
un-correlated with G25.1$-$2.3 (see Fig.~\ref{G25}). We determined
that the integrated flux density of G25.1$-$2.3 is $S_{6cm}$ =
3.7$\pm$0.4~Jy, $S_{11cm}$ = 4.7$\pm$0.5~Jy, and $S_{21cm}$ =
6.7$\pm$0.7~Jy, respectively. The spectrum that we fitted to these
data has a spectral index of $\alpha = -0.49\pm0.13$
(Fig.~\ref{spe}). SNR G25.1$-$2.3 has a surface brightness of $\rm
\Sigma_{1~GHz} = 5.0 \times 10^{-22}Wm^{-2}Hz^{-1}sr^{-1}$, which
makes it one of the fainter SNRs in Green's sample (\cite{Green09},
see his Fig.~1).

From a possibly associated cavity found in the neutral atomic gas, we
estimate a distance of 3.1~kpc to the new SNR G25.1$-$2.3. If this is
true, the distance of G25.1$-$2.3 can explain the absence of polarized
emission coming from this object, since polarized emission originated
beyond 3~kpc might not be detected at $\lambda$6\ cm in this direction
of the Galactic plane (\cite{Sun11}).

\section{Other SNRs discovered with the $\lambda$6\ cm data}

Based on high angular resolution synthesis observations,
\cite{Foster13} recently identified two other new SNRs, G152.4$-$2.1
and G190.9$-$2.2, which are even fainter than G156.2+5.7. The
$\lambda$6\ cm total intensity and polarization data from the
Sino-German $\lambda$6\ cm survey were incorporated in their study and
provide strong support and evidence for the identifications.

\begin{acknowledgements}
  We are grateful for financial support from the National Natural
  Science Foundation of China (10473015, 10773016) for the Sino-German
  $\lambda$6\ cm polarization survey of the Galactic plane. The
  Sino-German cooperation was supported via partner group of the MPIfR
  at the NAOC in the frame of the exchange program between the MPG and
  the CAS for many biliteral visits.
\end{acknowledgements}

\end{document}